\documentclass[prb,preprint,eqsecnum]{revtex4}%
\usepackage{amsmath}
\usepackage{graphicx}
\usepackage{amssymb}
\usepackage{amsfonts}%
\ifx\pdfoutput\relax\let\pdfoutput=\undefined\fi
\newcount\msipdfoutput
\ifx\pdfoutput\undefined\else
\ifcase\pdfoutput\else
\ifx\paperwidth\undefined\else
\ifdim\paperheight=0pt\relax\else\pdfpageheight\paperheight\fi
\ifdim\paperwidth=0pt\relax\else\pdfpagewidth\paperwidth\fi
\fi\fi\fi
\begin{document}
\title{Rapid evaluation of the periodic Green's function in $d$ dimensions}
\author{Sandeep Tyagi}
\email{s.tyagi@fias.uni-frankfurt.de}
\affiliation{Frankfurt Institute for Advanced Studies, J. W. Goethe Universit\"{a}t,
D-64038 Frankfurt am Main, Germany}

\begin{abstract}
A method is given to obtain the Green's function for the Poisson equation in
any arbitrary integer dimension under periodic boundary conditions. We obtain
recursion relations which relate the solution in $d$-dimensional space to that
in $\left(  d-1\right)  $-dimensional space. Near the origin, the Green's
function is shown to split in two parts, one is the essential Coulomb
singularity and the other part is regular. We are thus able to give
representations of the Coulomb sum in higher dimensions without taking
recourse to any integral representations. The expressions converge
exponentially fast in all part of the simulation cell. Works of several
authors are shown to be special cases of this more general method.

\end{abstract}
\maketitle

\section{Introduction}

The Poisson equation is probably one of the most useful equations in physics.
In a two-dimensional (2D) space, the periodic solution of this equation
corresponds to the solution of particles interacting with the logarithmic
interaction, and it has applications in simulations of 2D pancake vortices in
high-temperature superconductors \cite{tyagilog}. In 3D, periodic solutions to
the Poisson equation are used in electromagnetism.  Here, the solution of the
Poisson equation corresponds to a number of charges interacting with the
Coulomb potential. This 3D periodic solution is routinely used in most
simulations involving charged particles. Recently, the periodic solution of
the Poisson equation in higher dimensions has found use in the string theory.

In 1D and 2D, the Green's function for the Poisson equation for a charge
neutral box may be obtained in a closed form. In 3D, one can obtain rapidly
converging series representations using well known method by Ewald
\cite{ewald}. The other two approaches for the 3D case were given Lekner
\cite{lekner} and Sperb \cite{sperb}.  However, in higher dimensions, one can
either use the Ewald method which has its drawbacks, or use the Jacobi theta
function identities\cite{glasser}. In general, there is no efficient way to
calculate the Green's function in a general $d$-dimensional space with $d>3.$

In this paper, we give an exponentially fast converging series representation
for the Green's function of the Poisson equation in any positive integer
dimension. This work will generalize the methods employed for 2D and 3D case
\cite{tyagipre}, and will tie together the different approaches taken by
Lekner \cite{lekner} and Sperb \cite{sperb} for the especial case of $d=3$.
The outline of the paper is as follows. In Sec. I we derive expressions giving
the Coulomb sum in the $d$-dimensional space. In Sec. II we derive recursive
relations using the result of the previous section. In Sec. III we discuss the results.

\section{Green's function in $d$ dimensions}

For simplicity, we consider the case of a unit charge situated within a cubic
box in $d$ dimensions. The sides of the box are all assumed to be of unit
length. From here onwards, we will refer to the box as simulation cell. The
basic simulation cell repeats itself in all $d$ dimensions. We also assume a
charge neutral system. The unit charge interacts with other identical unit
charges (for the case of different charges $q_{1}$ and $q_{2}$ one just gets
an extra factor of $q_{1}q_{2}$) situated at the vertices of the periodic
structure. The periodic Green's function in $d$ dimensions satisfies the
Poisson equation,
\begin{equation}
\nabla_{d}^{2}G(\boldsymbol{r})=-C_{d}\,\sum_{\boldsymbol{l}}\delta
(\boldsymbol{r}\mathbf{+}\boldsymbol{l}),\label{2dim}%
\end{equation}
where $\nabla_{d}^{2}$ is the Laplacian operator in $d$ dimensions,
$\boldsymbol{l}$ denotes a $d$-dimensional vector, whose components are
integers ranging over $-\infty$ to $+\infty,$ and $C_{d}$ is a
dimension-dependent factor. The value of $C_{d}$ for various dimensions is%

\[
C_{d}=\left\{
\begin{array}
[c]{c}%
2\ \ \ \ \ \ \ \ \ \ \ \ \ \ \ \ \text{~\ }d=1\\
2\pi\text{\ \ \ \ \ \ \ \ \ \ \ \ \ \ \ \ \ }d=2,\\
4\pi^{\nu+1}/\Gamma\left(  \nu\right)  \text{ \ \ \ }d>2.
\end{array}
\right.
\]
Here, $\Gamma(\nu)$ stands for the Gamma function, and $\nu=\left(
d-2\right)  /2$. \ We note that with this choice of $C_{d}$ in Eq.(\ref{2dim}%
), the $G$ stands for the Coulomb type summation in $d$ dimensions. Thus, $G$
corresponds to a sum of type $-\left\vert \boldsymbol{r}\right\vert $ in 1D, a
logarithmic sum, -$\ln\left\vert \boldsymbol{r}\right\vert $, in 2D and a sum
of type $\left\vert \boldsymbol{r}\right\vert ^{-(d-2)}$ for a $d$-dimensional
space with $d>2$ . The solution of Eq. (\ref{2dim}) diverges, which is a
simple consequence of the fact that the interaction energy of a charge with
another charge and all its periodic images is infinite. To obtain a meaningful
value of $G$ we will have to modify Eq. (\ref{2dim}) as follows\cite{marshall}%
:%
\begin{equation}
\nabla^{2}G_{d}(\boldsymbol{r})=-C_{d}\,\sum_{\boldsymbol{l}}\delta
(\boldsymbol{r}\mathbf{+}\boldsymbol{l})+\frac{C_{d}}{l_{1}l_{2}\cdots l_{d}%
}.\label{22dim}%
\end{equation}
The second term in eq.(\ref{22dim}) amounts to the presence of a uniform
background charge. Thus, for every charge, $q$, one may imagine a uniform
distribution of charge, such that the total charge per basic simulation cell
adds up to $-q$. For a charge neutral periodic system, imposing these kind of
background uniform charge distributions does not matter since the total
uniform background charge adds up to zero. However, now a unit charge located
within the basic simulation cell at position $\left\{  x_{i}\right\}  $ not
only interacts with a second charge located at the origin and its periodic
images, but also interacts with the neutralizing background charge of the
second particle. This particular way of introducing the artificial
neutralizing background charge leads to only the intrinsic part \cite{lekner}
of the potential energy. We note that once the Green's function is obtained,
the solution of the equation
\[
\nabla^{2}V_{d}=-C_{d}\boldsymbol{\rho}\left(  \boldsymbol{r}\right)
\]
under periodic boundary conditions could be simply obtained from%
\[
V_{d}=\int_{\text{cell}}G_{d}\left(  \boldsymbol{r-r}^{\prime}\right)
\boldsymbol{\rho}\left(  \boldsymbol{r}^{\prime}\right)  d\boldsymbol{r}%
^{\prime},
\]
where $\boldsymbol{\rho}$ is periodic and the simulation cell is overall
charge neutral. \ The rapid evaluation of the $G_{d}$ is discussed in the next section.

The solution of eq.(\ref{22dim}) can be written easily in the Fourier space as
\cite{tyagipre} :%

\begin{align}
G_{d}(x_{1},x_{2},\cdots,x_{d})  &  =\frac{C_{d}}{(2\pi)^{2}}\times\nonumber\\
&  \lim_{\beta\rightarrow0}\left(  \sum_{\{m\}_{d}}\frac{e^{i2\pi(m_{1}%
x_{1}+m_{2}x_{2}+\cdots+m_{d}x_{d})}}{\left\{  m_{1}^{2}+m_{2}^{2}%
+\cdots+m_{d}^{2}+\beta^{2}/4\pi^{2}\right\}  }-\frac{4\pi^{2}}{\beta^{2}%
}\right)  , \label{g1}%
\end{align}
where $\beta$ is an infinitesimal parameter which tends to zero. Here , the
set $\{m_{1,d}\}$ denotes a set of $d$ integers $\{m_{1},m_{2},...,m_{d}\}$.
Each one of these integers $m_{i}$ runs over $-\infty$ to $+\infty.$ Also,
$x_{1},x_{2},...,x_{d}$ denote the components of vector $\boldsymbol{r}_{d}$
in $d$-dimensions. Due to the periodic boundary conditions, it is sufficient
to treat the case where each $x_{i}$ satisfies $-0.5<x\leq0.5$. The complete
expression for the potential has a term arising from the surface contribution.
For the 2D case this term turns out to be zero, but for 3D one obtains a
contribution from a dipole term \cite{deleeuw}. At this point, we would recast
the Eq. (\ref{g1}) in an alternative form. For that, we use the fact that the
solution of
\begin{equation}
\left(  \nabla^{2}-\beta^{2}\right)  Q_{0}(\boldsymbol{r})=-\delta
(\boldsymbol{r}) \label{q0}%
\end{equation}
in $d$-dimensional space is given by%
\begin{equation}
Q_{0}(\left\vert \boldsymbol{r}\right\vert ;\beta)=\frac{1}{\left(
2\pi\right)  ^{\nu+1}}\frac{\beta^{\nu}K_{\nu}\left(  \beta\left\vert
\boldsymbol{r}\right\vert \right)  }{r^{\nu}}.
\end{equation}
Thus, the solution of
\begin{equation}
\left(  \nabla^{2}-\xi^{2}\right)  Q_{d}(\boldsymbol{r};\beta)=-C_{d}%
\sum_{\boldsymbol{l}}\delta(\boldsymbol{r}\mathbf{+}\boldsymbol{l})
\end{equation}
in $d$-dimensional space will be given by%
\begin{equation}
Q_{d}(\boldsymbol{r};\xi)=\frac{C_{d}}{\left(  2\pi\right)  ^{\nu+1}}%
\sum_{\left\{  m_{1,d}\right\}  }\left[  \xi^{\nu}\frac{K_{\nu}\left(  \xi
r_{1,d}\right)  }{r_{1,d}^{\nu}}\right]  , \label{a14}%
\end{equation}
where%
\begin{equation}
r_{1,d}=\left[  \sum_{i=1}^{d}\left(  m_{i}-x_{i}\right)  ^{2}\right]  ^{1/2}
\label{a11}%
\end{equation}
On the other hand, the solution of Eq. (\ref{q0}) can be written down in the
Fourier space easily as
\begin{equation}
Q_{d}(\boldsymbol{r};\beta)=\frac{C_{d}}{\left(  2\pi\right)  ^{2}}%
\sum_{\{m\}_{d}}\frac{e^{i2\pi(m_{1}x_{1}+m_{2}x_{2}+\cdots+m_{d}x_{d})}%
}{\left\{  m_{1}^{2}+m_{2}^{2}+\cdots+m_{d}^{2}+\beta^{2}/4\pi^{2}\right\}  }.
\end{equation}
Using Eqs.(\ref{g1}) and (\ref{q0}) we see that one can write%
\begin{equation}
G_{d}(x_{1},x_{2},\cdots,x_{d})=C_{d}\lim_{\beta\rightarrow0}\left(  \frac
{1}{\left(  2\pi\right)  ^{\nu+1}}\sum_{\left\{  m_{1,d}\right\}  }\left[
\beta^{\nu}\frac{K_{\nu}\left(  \beta r_{1,d}\right)  }{r_{1,d}^{\nu}}\right]
-\frac{1}{\beta^{2}}\right)  . \label{ak}%
\end{equation}
A yet another alternative form of $G_{d}$ can be obtained as follows. We can
perform one of the $d$ sums in Eq.(\ref{g1}) analytically using the formula
\cite{gradshteyn}
\begin{equation}
\sum_{i=-\infty}^{\infty}\frac{\exp\left(  2\pi imx\right)  }{m^{2}+\gamma
^{2}}=\frac{\pi}{\gamma}\frac{\cosh\left[  \pi\gamma\left(  1-2\left\vert
x\right\vert \right)  \right]  }{\sinh\left(  \pi\gamma\right)  }.
\end{equation}
Thus, we obtain%
\begin{align}
G_{d}(x_{1},x_{2},\cdots,x_{d})  &  =\frac{C_{d}}{\left(  2\pi\right)  ^{2}%
}\lim_{\beta\rightarrow0}\left(  \sum_{\left\{  m_{2,d}\right\}  \ }\frac{\pi
}{\gamma_{\left\{  m_{2,d}\right\}  }}\frac{\cosh\left[  \pi\gamma_{\left\{
m_{2,d}\right\}  }\left(  1-2\left\vert x_{1}\right\vert \right)  \right]
}{\sinh\left(  \pi\gamma_{\left\{  m_{2,d}\right\}  }\right)  }\right.
\nonumber\\
&  \left.  \times\exp\left[  2\pi i\sum_{i=2}^{d}m_{i}x_{i}\right]  -\frac
{1}{\beta^{2}}\right)  , \label{gd}%
\end{align}
where $\gamma_{\{m_{2,d}\}}$ is defined as
\begin{equation}
\gamma_{\left\{  m_{2,d}\right\}  }=\left(  \sum_{i=2}^{d}m_{i}^{2}+\beta
^{2}\right)  ^{1/2}. \label{a39}%
\end{equation}
For the purpose of taking the limit $\beta\rightarrow0,$ the sum in the first
part of Eq. (\ref{gd}) is broken as
\begin{equation}
\sum_{\left\{  m_{2,d}\right\}  }=\sum_{\left\{  m_{2,d}\right\}  }^{\prime
}+\left(  \text{Term with }m_{2}=0,m_{3=0}...,m_{d}=0\right)  , \label{a40}%
\end{equation}
where a prime over the summation sign indicates that the term corresponding to
all $m_{i}$ being zero is to be excluded from the summation. This leads to the
following representation for $G_{d}$:%
\begin{align}
G_{d}  &  =\frac{C_{d}}{\left(  2\pi\right)  ^{2}}\sum_{\left\{
m_{2,d}\right\}  }^{\prime}\frac{\pi}{\gamma_{\left\{  m_{2,d}\right\}  }%
}\frac{\cosh\left[  \pi\gamma_{\left\{  m_{2,d}\right\}  }\left(
1-2\left\vert x_{1}\right\vert \right)  \right]  }{\sinh\left(  \pi
\gamma_{\left\{  m_{2,d}\right\}  }\right)  }\nonumber\\
&  \times\exp\left(  2\pi i\sum_{i=2}^{d}m_{i}x_{i}\right)  +H_{d},
\label{hkm}%
\end{align}
where we have taken the limit $\beta\rightarrow0,$ i.e. we have substituted
$\beta=0$ in the first part, and $H_{d}$ is given by%
\begin{align}
H_{d}  &  =\frac{C_{d}}{\left(  2\pi\right)  ^{2}}\lim_{\beta\rightarrow
0}\left(  \frac{2\pi^{2}}{\beta}\frac{\cosh\left[  \left(  1/2-\left\vert
x_{1}\right\vert \right)  \beta\right]  }{\sinh\left(  \beta/2\right)  }%
-\frac{4\pi^{2}}{\beta^{2}}\right) \nonumber\\
&  =C_{d}\frac{1}{12}\left(  1-6\left\vert x_{1}\right\vert +6x_{1}%
^{2}\right)  . \label{a41}%
\end{align}
To avoid the bad convergence towards $x_{1}\rightarrow0,$ we further modify
the summation in the first part of Eq. (\ref{hkm})\ by using the following
trigonometric identity
\begin{equation}
\frac{\cosh(a-b)}{\sinh\left(  b\right)  }=\exp(-{b})\frac{\cosh(a)}{\sinh
(b)}+\exp(-a). \label{a42}%
\end{equation}
Thus, $G_{d}$ can be written as
\begin{equation}
G_{d}=H_{d}+J_{d}+M_{d}, \label{mkh}%
\end{equation}
where $H_{d}$ is defined in Eq. (\ref{a41}), $J_{d}$ is given by%
\begin{align}
J_{d}  &  =\frac{C_{d}}{\left(  2\pi\right)  ^{2}}\sum_{\left\{
m_{2,d}\right\}  }^{\prime}\frac{\pi}{\gamma_{\left\{  m_{2,d}\right\}  }}%
\exp\left(  -\pi\gamma_{\left\{  m_{2,d}\right\}  }\right) \nonumber\\
&  \times\frac{\cosh\left[  \pi\gamma_{\left\{  m_{2,d}\right\}  }\left(
1-2\left\vert x_{1}\right\vert \right)  \right]  }{\sinh\left(  \pi
\gamma_{\left\{  m_{2,d}\right\}  }\right)  }\exp\left(  2\pi i\sum_{i=2}%
^{d}m_{i}x_{i}\right)  , \label{a43}%
\end{align}
and%
\begin{equation}
M_{d}=\frac{C_{d}}{\left(  2\pi\right)  ^{2}}\sum_{\left\{  m_{2,d}\right\}
}^{\prime}\frac{\pi}{\gamma_{\left\{  m_{2,d}\right\}  }}\exp\left[
-2\left\vert x_{1}\right\vert \pi\gamma_{\left\{  m_{2,d}\right\}  }\right]
\exp\left(  2\pi i\sum_{i=2}^{d}m_{i}x_{i}\right)  . \label{a44}%
\end{equation}
It is easy to see that Eq. (\ref{a43}) does not have any convergence problem
as $x_{1}$tends to zero. Thus, the whole problem has reduced to evaluating the
$M_{d}$ term efficiently. This will be done in the next section.

\section{Recursive Formulas}

In this section we obtain recursive formulas for $G_{d}$ in two different
ways, starting with the expressions in Eq.\ (\ref{ak})\ and (\ref{mkh})
respectively. The first method, with Eq. (\ref{ak}) as the starting point,
will contain Lekner's results for $d=3$ as a special case, while the second
method will contain Sperb's result in 3D as a special case. With the help of
Eqs. (\ref{ak}) and (\ref{g1}) we can write%
\begin{align}
G_{d}(x_{1},x_{2},...,x_{d}) &  =\frac{C_{d}}{\left(  2\pi\right)  ^{2}}%
\lim_{\beta\rightarrow0}\left(  Q_{d}(x_{1},x_{2},..,x_{d};\beta)-\frac
{1}{\beta^{2}}\right)  \nonumber\\
&  =\frac{C_{d}}{\left(  2\pi\right)  ^{2}}\lim_{\beta\rightarrow0}\left(
\sum_{\left\{  m_{1,d}\right\}  }\exp\left(  2\pi im_{1}x\right)  \times
\frac{\exp\left(  2\pi i\sum_{i=2}^{d}m_{i}x_{i}\right)  }{\sum_{i=2}^{d}%
m_{i}^{2}+\left[  \beta^{2}+m_{1}^{2}\right]  }\right)  -\frac{1}{\beta^{2}%
}.\label{a22}%
\end{align}
Using the definition of $G_{d}$, Eq. (\ref{a22}) be written as
\begin{align}
G_{d}(x_{1},x_{2},...,x_{d}) &  =\frac{C_{d}}{C_{d-1}}\lim_{\beta\rightarrow
0}\left[  \sum_{m_{1}}\exp\left(  2\pi im_{1}x_{1}\right)  \right.
\nonumber\\
&  \left.  \times Q_{d-1}\left(  x_{2},..x_{d};\sqrt{\beta^{2}+\left(  2\pi
m_{1}\right)  ^{2}}\right)  -\frac{C_{d-1}}{\beta^{2}}\right]  .\label{b1}%
\end{align}
We separate out the term corresponding to $m_{1}=0$ in Eq. (\ref{b1}) so that
the limit corresponding to $\beta$ can be taken. Thus, we write Eq.
(\ref{a22}) as%
\begin{align}
G_{d}(x_{1},x_{2},...,x_{d}) &  =2\frac{C_{d}}{C_{d-1}}\sum_{m_{1}=1}^{\infty
}\cos\left(  2\pi m_{1}x_{1}\right)  Q_{d-1}\left(  x_{2},x_{3},...,x_{d};2\pi
m_{1}\right)  \nonumber\\
&  +\frac{C_{d}}{C_{d-1}}\lim_{\beta\rightarrow0}\left[  Q_{d-1}\left(
x_{2},x_{3},...,x_{d};\beta\right)  -\frac{C_{d-1}}{\beta^{2}}\right]
\nonumber\\
&  =2\frac{C_{d}}{C_{d-1}}\sum_{m_{1}=1}^{\infty}\cos\left(  2\pi m_{1}%
x_{1}\right)  Q_{d-1}\left(  x_{2},x_{3},...,x_{d};2\pi m_{1}\right)
\nonumber\\
&  +\frac{C_{d}}{C_{d-1}}G_{d-1}(x_{2},..,x_{d}),\label{a24}%
\end{align}
where we have taken the limit $\beta\rightarrow0$ in the first term. The Eq.
(\ref{a24}) is one of the most important result of this paper. This relates a
$d$-dimensional sum to a $\left(  d-1\right)  $-dimensional sum. This is a
recursive relation. If one is able to obtain the Green function for the
$\left(  d-1\right)  $-dimensional space, one can obtain the Green's function
for the $d$-dimensional space. The first term in Eq. (\ref{a24}) can be
modified in the following way. We can use a form of $G_{d-1}$ similar to the
one used in Eq. (\ref{a14}) to obtain%
\begin{align}
G_{d}(x_{1},x_{2},...,x_{d}) &  =2\frac{C_{d}}{\left(  2\pi\right)  ^{\nu
+1/2}}\sum_{m_{1}=1}^{\infty}\sum_{\left\{  m_{2,d}\right\}  }\cos\left(  2\pi
m_{1}x_{1}\right)  \nonumber\\
&  \times\left(  2\pi m_{1}\right)  ^{\nu-1/2}\frac{K_{\nu-1/2}\left(  2\pi
m_{1}r_{2,d}\right)  }{r_{2,d}^{\nu-1/2}}+\frac{C_{d}}{C_{d-1}}G_{d-1}\left(
x_{2},..,x_{d}\right)  ,\label{a25}%
\end{align}
where $\{m_{2,d}\}$ denotes a sum over sets $\{m_{2,}m_{3},...,m_{d}\}$ and
$r_{2,d}$ is defined like Eq. (\ref{a11})%
\begin{equation}
r_{2,d}=\left[  \sum_{i=2}^{d}\left(  m_{i}-x_{i}\right)  ^{2}\right]
^{1/2}.\label{a26}%
\end{equation}
Let us now consider three different cases corresponding to $d=1,$ $d=2$ and
$d>2.$ For $d=1$ we can evaluate $G_{d=1}$ in a closed form:
\begin{align}
G_{1} &  =\frac{C_{1}}{\left(  2\pi\right)  ^{2}}\lim_{\beta\rightarrow
0}\left(  \sum_{m_{1}}\frac{\exp\left(  2\pi im_{1}x_{1}\right)  }{\beta
^{2}+m_{1}^{2}}-\frac{1}{\beta^{2}}\right)  \nonumber\\
&  =\frac{C_{1}}{\left(  2\pi\right)  ^{2}}\lim_{\beta\rightarrow0}\left(
\frac{\pi}{\beta}\frac{\cosh\left[  \pi\beta\left(  1-2\left\vert
x_{1}\right\vert \right)  \right]  }{\sinh\left[  \pi\beta\right]  }-\frac
{1}{\beta^{2}}\right)  \nonumber\\
&  =C_{1}\frac{1}{12}\left(  1-6\left\vert x_{1}\right\vert +6x_{1}%
^{2}\right)  .
\end{align}
Also, the self-energy for this case may be obtained as%
\[
G_{1}^{\text{self}}=\lim_{x_{1}\rightarrow0}G_{1}+\left\vert x_{1}\right\vert
=\frac{C_{1}}{12}.
\]
For $d=2$ case, we obtain using Eq. (\ref{ak}):%
\begin{align}
G_{2}(x_{1},x_{2}) &  =2\frac{C_{2}}{\left(  2\pi\right)  ^{1/2}}\sum
_{m_{1}=1}^{\infty}\sum_{m_{2}=-\infty}^{+\infty}\cos\left(  2\pi m_{1}%
x_{1}\right)  \left(  2\pi m_{1}\right)  ^{-1/2}\nonumber\\
&  \times\frac{K_{-1/2}\left(  2\pi m_{1}\left\vert x_{2}+m_{2}\right\vert
\right)  }{\left\vert x_{2}+m_{2}\right\vert ^{-1/2}}+\frac{C_{2}}{C_{1}}%
G_{1}\left(  x_{2}\right)  .
\end{align}
Now, using the relation \cite{gradshteyn} ,
\begin{equation}
K_{-1/2}\left(  r\right)  =\sqrt{\frac{\pi}{2r}}\exp\left(  -r\right)  ,
\end{equation}
we can write%
\begin{align}
G_{2}(x_{1},x_{2}) &  =\frac{C_{2}}{2\pi}\sum_{m_{2}=-\infty}^{+\infty}%
\sum_{m_{1}=1}^{\infty}\frac{\cos\left(  2\pi m_{1}x_{1}\right)  }{\left\vert
m_{1}\right\vert }\nonumber\\
&  \times\exp\left(  -2\pi m_{1}\left\vert x_{2}+m_{2}\right\vert \right)
+\frac{C_{2}}{C_{1}}G_{1}\left(  x_{2}\right)  .
\end{align}
The sum over $m_{1}$ can be easily carried out using the identity
\cite{tyagipre}%
\begin{align}
L(x_{1},x_{2}) &  =\sum_{m_{1}=1}^{\infty}\frac{\cos\left(  2\pi m_{1}%
x_{1}\right)  }{m_{1}}\exp\left(  -2\pi m_{1}\left\vert x_{2}\right\vert
\right)  \nonumber\\
&  =-\frac{1}{2}\ln\left(  1-2\exp\left[  -2\pi x_{2}\right]  \cos\left[  2\pi
x_{1}\right]  +\exp\left[  -4\pi x_{2}\right]  \right)  .\label{lx}%
\end{align}
Thus, $G_{2}$ can be written as%
\begin{align}
G_{2}(x_{1},x_{2}) &  =\frac{C_{2}}{2\pi}\sum_{m_{2}=1}^{+\infty}%
L(x_{1},\left\vert x_{2}+m_{2}\right\vert )+L(x_{1},\left\vert x_{2}%
-m_{2}\right\vert )\nonumber\\
&  +L\left(  x_{1},x_{2}\right)  +\frac{C_{2}}{C_{1}}G_{1}\left(
x_{2}\right)  .\label{g2e}%
\end{align}
It is also trivial to derive
\begin{equation}
G_{2}^{\text{self}}=2\frac{C_{2}}{2\pi}\sum_{m_{2}=1}^{+\infty}L(0,\left\vert
m_{2}\right\vert )-\ln2\pi+\frac{C_{2}}{12}.\label{self1}%
\end{equation}
\newline Now we consider the case for $d>2.$ We can obtain $G_{d}$ from Eq.
(\ref{a25}). It is seen that for large arguments the modified Bessel functions
decay as%
\begin{equation}
K_{\nu}\left(  r\right)  \sim\sqrt{\frac{\pi}{2r}}\exp\left(  -r\right)
.\label{a27}%
\end{equation}
As a result, the first term in Eq. (\ref{a25}) decays exponentially. However,
one may run into problem if $r_{2,d\text{ }}$is very small. In such a case the
terms corresponding to $\{m_{2,d}\}$ all being zero form a very slowly
converging series over $m_{1}.$ This problem of slow convergence when
$r_{2,d}$ is small can be handled in the following recursive manner. We
separate out the particular terms corresponding to $\{m_{2,d}\}$ all being
zero, and define
\begin{align}
E_{d}(x_{1},x_{2},...,x_{d}) &  =2\frac{C_{d}}{\left(  2\pi\right)  ^{\nu
+1/2}}\sum_{m_{1}=1}^{\infty}\sum_{\left\{  m_{2,d}\right\}  }\cos\left(  2\pi
m_{1}x_{1}\right)  \nonumber\\
&  \times\left(  2\pi m_{1}\right)  ^{\nu-1/2}\frac{K_{\nu-1/2}\left(  2\pi
m_{1}r_{2,d}\right)  }{r_{2,d}^{\nu-1/2}}\nonumber\\
&  =2\frac{C_{d}}{\left(  2\pi\right)  ^{\nu+1/2}}\sum_{m_{1}=1}^{\infty}%
\sum_{\left\{  m_{2,d}\right\}  }^{\prime}\cos\left(  2\pi m_{1}x_{1}\right)
\nonumber\\
&  \times\left(  2\pi m_{1}\right)  ^{\nu-1/2}\frac{K_{\nu-1/2}\left(  2\pi
m_{1}r_{2,d}\right)  }{r_{2,d}^{\nu-1/2}}\nonumber\\
&  +2\frac{C_{d}}{\left(  2\pi\right)  ^{\nu+1/2}}\sum_{m_{1}=1}^{\infty}%
\cos\left(  2\pi m_{1}x_{1}\right)  \nonumber\\
&  \times\left(  2\pi m_{1}\right)  ^{\nu-1/2}\frac{K_{\nu-1/2}\left(  2\pi
m_{1}r\right)  }{r^{\nu-1/2}}.
\end{align}
Now, we show how to handle the evaluation of $E_{d}$ corresponding to $d>3.$
The case for $d=3$ will be almost the same. Using the relation
\cite{gradshteyn} (which by the way can be derived from Eq. (\ref{a22}))
\begin{align}
\sum_{k=-\infty}^{\infty}\frac{1}{\left[  \left(  x+k\right)  ^{2}%
+r^{2}\right]  ^{\frac{1}{2}+\nu}} &  =\frac{\sqrt{\pi}}{\Gamma\left(
\nu+\frac{1}{2}\right)  }\left\{  \frac{\Gamma\left(  \nu\right)  }{r^{2\nu}%
}+4\left(  \frac{\pi}{r}\right)  ^{\nu}\right.  \nonumber\\
&  \left.  \times\sum_{l=1}^{\infty}l^{\nu}K_{\nu}\left(  2\pi lr\right)
\cos\left(  2\pi lx\right)  \right\}  \ \ \ \ \ \ \ \ \ \nu>0,\label{a29}%
\end{align}
we can write%
\begin{align}
E_{d}(x_{1},x_{2},...,x_{d}) &  =2\frac{C_{d}}{\left(  2\pi\right)  ^{\nu
+1/2}}\sum_{m_{1}=1}^{\infty}\sum_{\left\{  m_{2,d}\right\}  }^{\prime}%
\cos\left(  2\pi m_{1}x_{1}\right)  \nonumber\\
&  \times\left[  \left(  2\pi m_{1}\right)  ^{\nu-1/2}\frac{K_{\nu-1/2}\left(
2\pi m_{1}r_{2,d}\right)  }{r_{2,d}^{\nu-1/2}}\right]  \nonumber\\
&  +\sum_{k=-\infty}^{\infty}\frac{1}{\left[  \left(  x+k\right)  ^{2}%
+r^{2}\right]  ^{\nu}}-\frac{\sqrt{\pi}}{\Gamma\left(  \nu\right)  }%
\frac{\Gamma\left(  \nu-1/2\right)  }{r^{2\nu-1}}.\label{a30}%
\end{align}
Also, the sum over $k$ in Eq. (\ref{a30}) can be written as
\begin{align}
\sum_{k=-\infty}^{\infty}\frac{1}{\left[  \left(  x+k\right)  ^{2}%
+r^{2}\right]  ^{\nu}} &  =\frac{1}{\left(  x^{2}+r^{2}\right)  ^{\nu}}%
+\sum_{k=1}^{N-1}\left(  \frac{1}{\left[  \left(  x+k\right)  ^{2}%
+r^{2}\right]  ^{\nu}}+\frac{1}{\left[  \left(  x-k\right)  ^{2}+r^{2}\right]
^{\nu}}\right)  \nonumber\\
&  +\sum_{l=1}^{\infty}\binom{-\nu}{l}r^{2l}\left[  \zeta\left(
2l+2\nu,N+x\right)  +\zeta\left(  2l+2\nu,N-x\right)  \right]  ,\label{a31}%
\end{align}
where $N$ is an arbitrary integer\cite{tyagipre} such that $N>r+\left\vert
x\right\vert $. Using Eqs. (\ref{a25})\ , (\ref{a30}) and (\ref{a31}) we can
now write
\begin{align}
G_{d}(x_{1},x_{2},..x_{d})-\frac{1}{\left(  x_{1}^{2}+r_{\perp}^{2}\right)
^{\nu+1/2}} &  =2\frac{C_{d}}{\left(  2\pi\right)  ^{\nu+1/2}}\sum_{m_{1}%
=1}^{\infty}\sum_{\left\{  m_{2,d}\right\}  }^{\prime}\cos\left(  2\pi
m_{1}x_{1}\right)  \nonumber\\
&  \times\left(  2\pi m_{1}\right)  ^{\nu-1/2}\frac{K_{\nu-1/2}\left(  2\pi
m_{1}r_{2,d}\right)  }{r_{2,d}^{\nu-1/2}}\nonumber\\
&  +\sum_{k=1}^{N-1}\left(  \frac{1}{\left[  \left(  x_{1}+k\right)
^{2}+r_{\perp}^{2}\right]  ^{\nu}}+\frac{1}{\left[  \left(  x_{1}-k\right)
^{2}+r_{\perp}^{2}\right]  ^{\nu}}\right)  \nonumber\\
&  +\sum_{l=1}^{\infty}\binom{-\nu}{l}r_{\perp}^{2l}\left[  \zeta\left(
2l+2\nu,N+x_{1}\right)  +\zeta\left(  2l+2\nu,N-x_{1}\right)  \right]
\nonumber\\
&  +\frac{C_{d}}{C_{d-1}}\left(  G_{d}(x_{2},..x_{d})-\frac{1}{r_{\perp}%
^{2\nu}}\right)  .\label{a32}%
\end{align}
Note that if $d=3$ then instead of Eq.(\ref{a31}) we should use%
\begin{align}
&  4\sum_{m_{1}=1}^{\infty}\,K_{0}\left(  2\pi m_{1}\left(  x_{2}^{2}%
+x_{3}^{2}\right)  ^{1/2}\right)  \cos\left(  2\pi m_{1}x_{1}\right)
\nonumber\\
&  =2\left\{  \gamma+\ln\left(  \frac{\left(  x_{2}^{2}+x_{3}^{2}\right)
^{1/2}}{2}\right)  \right\}  +\frac{1}{\sqrt{x_{1}^{2}+x_{2}^{2}+x_{3}^{2}}%
}+S\left(  x_{1},x_{2},x_{3}\right)  ,\label{iden2}%
\end{align}
where%
\begin{align}
S\left(  x_{1},x_{2},x_{3}\right)   &  =\sum_{n=1}^{N-1}\left(  \frac{1}%
{\sqrt{x_{2}^{2}+x_{3}^{2}+\left(  n+x_{1}\right)  ^{2}}}+\frac{1}{\sqrt
{x_{2}^{2}+x_{3}^{2}+\left(  n-x_{1}\right)  ^{2}}}\right)  \nonumber\\
&  +\frac{1}{\sqrt{x_{1}^{2}+x_{2}^{2}+x_{3}^{2}}}-2\gamma-\left[
\psi(N+x_{1})+\psi(N-x_{1})\right]  \nonumber\\
&  +\sum_{l=1}^{\infty}\binom{-1/2}{l}\left(  x_{2}^{2}+x_{3}^{2}\right)
^{l}\left[  \zeta\left(  2l+1,N+x\right)  +\zeta\left(  2l+1,N-x\right)
\right]  .
\end{align}
Thus, for the 3D case one would make the following two changes in the
expression given in Eq. ( \ref{a32}). First, there would be an extra term
containing $-2\gamma-\left[  \psi(N+x_{1})+\psi(N-x_{1})\right]  $ on the
right hand side, and second the last term in Eq. (\ref{a32}) would be changed
to
\begin{equation}
\frac{C_{d}}{C_{d-1}}\left[  G_{d-1}\left(  x_{2},x_{3}\right)  +\ln r_{\perp
}\right]  .
\end{equation}
Eq. (\ref{a32}) provides us with a general algorithm to calculate $G_{d}$
efficiently in any dimensions. For an example, if we had started out with
$d=10$, we can obtain $G_{10}-r_{8}^{-8}$ by calculating $G_{9}-r_{7}^{-7}.$
Continuing in this fashion we will come down to calculating $G_{2}+\ln r_{2}.$
Now, this last part $G_{2}+\ln r_{2}$ has been obtained by several authors .
In fact, it can be obtained in a closed form \cite{glasser}. Thus, we have
been able to calculate $G_{10}-r_{8}^{-8}$ from which can obtain $G_{10}$ by
taking the radial part $r_{8}^{-8}$ on the other side. Other forms of $G_{2}$
are given by Gr$\;\not o\;$nbech-Jensen \cite{niels} and Tyagi \cite{tyagipre}%
. For the sake of completion we write down the result for $G_{2}$:%
\begin{align}
G_{2}(x_{1},x_{2}) &  =\frac{1}{2\pi}\sum_{m}^{\prime}\frac{\pi}{\left\vert
m\right\vert }\frac{\exp\left(  -\pi\left\vert m\right\vert \right)
\cosh\left[  2\pi mx_{1}\right]  }{\sinh\left(  \pi\left\vert m\right\vert
\right)  }\cos\left(  2\pi mx_{2}\right)  \nonumber\\
&  -\frac{1}{2}\ln\left[  \cosh\left(  2\pi x_{1}\right)  -\cos\left(  2\pi
x_{2}\right)  \right]  \nonumber\\
&  +\frac{\pi}{6}\left(  1+6x_{2}^{2}\right)  -\frac{\ln\left(  2\right)  }%
{2}.\label{a35}%
\end{align}
In the closed form $G_{2}$ is written as \cite{logclosed}%
\begin{equation}
G_{2}(x_{1},x_{2})=2\pi\left(  \frac{x_{2}^{2}}{2}-\frac{\ln2}{6\pi}+\frac
{1}{2\pi}\ln\left\vert \frac{\vartheta_{1}\left[  \pi\left(  x_{1}%
+ix_{2}\right)  ,\exp\left(  -\pi\right)  \right]  }{\vartheta_{1}^{^{\prime}%
}\left[  0,\exp\left(  -\pi\right)  \right]  ^{1/3}}\right\vert \right)
,\label{closed}%
\end{equation}
where $\vartheta_{1}$ represents the Jacobi theta function of the first kind.
Also, the self-energy for the 2D case can be obtained from Eq. (\ref{a35})%
\begin{equation}
G_{2}^{\text{self}}=\frac{1}{\pi}\sum_{m=1}^{\infty}\frac{\pi}{\left\vert
m\right\vert }\frac{\exp\left(  -\pi\left\vert m\right\vert \right)  }%
{\sinh\left(  \pi\left\vert m\right\vert \right)  }-\ln\left(  2\pi\right)
+\frac{\pi}{6},\label{self2}%
\end{equation}
or it can be obtained from Eq. (\ref{closed}):%
\begin{equation}
G_{2}^{\text{self}}=-\frac{\ln2}{3}-\ln\pi-\frac{2}{3}\ln\left\vert \left[
\vartheta_{1}^{^{\prime}}\left(  0,q\right)  \right]  \right\vert
.\label{self3}%
\end{equation}
All three forms Eq. (\ref{self1})\ , Eq. (\ref{self2})\ and Eq. (\ref{self3}%
)\ are equivalent and give numerically the same value for the self-energy.
Similarly Eqs. (\ref{g2e}), (\ref{a35}) and (\ref{closed}) show perfect agreement.

Now, we give another alternative approach. This time we start with Eq.
(\ref{mkh}), where $H_{d},$ $J_{d}$ and $M_{d}$ are defined in Eqs.
(\ref{a41}), (\ref{a43}) and (\ref{a44}). $H_{d}$ and $J_{d}$ do not have any
convergence problem in the region of interest. We show how to handle $M_{d}$.
A recursion formula similar to Eq. (\ref{a25}) can be established for $M_{d}$.
It is easy to see just by inspection that $M_{d}$ obeys the following
recursion formula:%
\begin{align}
M_{d}  &  =\frac{C_{d}}{C_{d-1}}M_{d-1}+2\frac{C_{d}}{\left(  2\pi\right)
^{\nu+1/2}}\sum_{m_{2}=1}^{\infty}\cos\left(  2\pi m_{2}x_{2}\right)  \left(
2\pi m_{2}\right)  ^{\nu-1/2}\nonumber\\
&  \times\sum_{\{m_{3,d}\}}\frac{K_{\nu-1/2}\left(  2\pi m_{2}\sqrt{x_{1}%
^{2}+\left(  m_{3}-x_{3}\right)  ^{2}+..+\left(  m_{d}-x_{d}\right)  ^{2}%
}\right)  }{\left[  \sqrt{x_{1}^{2}+\left(  m_{3}-x_{3}\right)  ^{2}%
+..+\left(  m_{d}-x_{d}\right)  ^{2}}\right]  ^{\nu-1/2}}, \label{md}%
\end{align}
where $M_{d-1}$, analogues to Eq. (\ref{a44}), stands for%
\begin{equation}
M_{d-1}=\frac{C_{d-1}}{\left(  2\pi\right)  ^{2}}\sum_{\left\{  m_{3,d}%
\right\}  }^{\prime}\frac{\pi}{\gamma_{\left\{  m_{3,d}\right\}  }}\exp\left[
-2\left\vert x_{1}\right\vert \pi\gamma_{\left\{  m_{3,d}\right\}  }\right]
\exp\left(  2\pi i\sum_{i=3}^{d}m_{i}x_{i}\right)  . \label{a47}%
\end{equation}
In the final step, we break the sum in the second part of Eq. (\ref{md}) as
follows%
\begin{equation}
\sum_{\{m_{3,d}\}}=\sum_{\{m_{3,d}\}}^{\prime}+\sum_{m_{3}=0,m_{4}=0,..}.
\label{a48}%
\end{equation}
The term corresponding to $m_{3}=0,$ $m_{4}=0...$ gives rise to a term $F_{d}$
in Eq. (\ref{md}):
\begin{align}
F_{d}  &  =2\frac{C_{d}}{\left(  2\pi\right)  ^{\nu+1/2}}\sum_{m_{2}%
=1}^{\infty}\cos\left(  2\pi m_{2}x_{2}\right)  \left(  2\pi m_{2}\right)
^{\nu-1/2}\nonumber\\
&  \times\frac{K_{\nu-1/2}\left(  2\pi m_{2}\sqrt{x_{1}^{2}+x_{3}^{2}%
+..+x_{d}^{2}}\right)  }{\left[  \sqrt{x_{1}^{2}+x_{3}^{2}+..+x_{d}^{2}%
}\right]  ^{\nu-1/2}}\nonumber\\
&  =\frac{1}{\sqrt{x_{1}^{2}+x_{2}^{2}+x_{3}^{2}+..+x_{d}^{2}}}-\frac{C_{d}%
}{C_{d-1}}\frac{1}{\sqrt{x_{1}^{2}+x_{3}^{2}+..+x_{d}^{2}}}\nonumber\\
&  +\sum_{k=1}^{N-1}\left(  \frac{1}{\left[  \left(  x_{2}+k\right)
^{2}+r^{2}\right]  ^{\nu}}+\frac{1}{\left[  \left(  x_{2}-k\right)  ^{2}%
+r^{2}\right]  ^{\nu}}\right) \nonumber\\
&  +\sum_{l=1}^{\infty}\binom{-\nu}{l}r^{2l}\left[  \zeta\left(
2l+2\nu,N+x_{2}\right)  +\zeta\left(  2l+2\nu,N-x_{2}\right)  \right]  .
\label{a49}%
\end{align}
Thus, we finally obtain the following recursion relationship for $M_{d}$:%
\begin{align}
\left(  M_{d}-\frac{1}{\left(  x_{1}^{2}+r_{\perp}^{2}\right)  ^{\nu+1/2}%
}\right)   &  =\frac{C_{d}}{C_{d-1}}\left(  M_{d-1}-\frac{1}{r_{\perp}^{2\nu}%
}\right) \nonumber\\
&  +2\frac{C_{d}}{\left(  2\pi\right)  ^{\nu+1/2}}\sum_{m_{2}=1}^{\infty}%
\cos\left(  2\pi m_{2}x_{2}\right)  \left(  2\pi m_{2}\right)  ^{\nu
-1/2}\nonumber\\
&  \times\sum_{\{m_{3,d}\}}^{\prime}\frac{K_{\nu-1/2}\left(  2\pi m_{2}%
\sqrt{x_{1}^{2}+\left(  m_{3}-x_{3}\right)  ^{2}+...+\left(  m_{d}%
-x_{d}\right)  ^{2}}\right)  }{\left[  \sqrt{x_{1}^{2}+\left(  m_{3}%
-x_{3}\right)  ^{2}+...+\left(  m_{d}-x_{d}\right)  ^{2}}\right]  ^{\nu-1/2}%
}\nonumber\\
&  +\sum_{k=1}^{N-1}\left(  \frac{1}{\left[  \left(  x_{2}+k\right)
^{2}+r_{\perp}^{2}\right]  ^{\nu}}+\frac{1}{\left[  \left(  x_{2}-k\right)
^{2}+r_{\perp}^{2}\right]  ^{\nu}}\right) \nonumber\\
&  +\sum_{l=1}^{\infty}\binom{-\nu}{l}r_{\perp}^{2l}\left[  \zeta\left(
2l+2\nu,N+x_{2}\right)  +\zeta\left(  2l+2\nu,N-x_{2}\right)  \right]  ,
\label{a50}%
\end{align}
where%
\begin{equation}
r_{\perp}^{2}=x_{1}^{2}+x_{3}^{2}+...+x_{d}^{2}.
\end{equation}
For $d=3$ case, once again, we will have to make two modifications in Eq.
(\ref{a50}). With this approach we have obtained Eq. (\ref{a50}), which is
analogues to Eq. (\ref{a32}). However, the analysis has become a little bit
tedious. The advantage of the second method is that it reduces the computation
time, as there is one less summation. The second advantage it can be written
down in a product decomposition form. For example, how such a product
decomposition form may be written, one may consult Sperb, where a special case
corresponding to $d=3$ is considered. In general, the procedure of dimensional
reduction is to be continued until we have $M_{1}$ on the left hand side. It
is clear that $M_{1}=0.$ Let us again consider three special cases. For $d=1$
one only has $H_{d=1}$ and thus $G_{1}=H_{1}.$ For $d=2$ one obtains%
\begin{align}
J_{2}  &  =\frac{C_{2}}{\left(  2\pi\right)  ^{2}}\sum_{m_{2}}^{\prime}%
\frac{\pi}{\gamma_{m_{2}}}\exp\left(  -\pi\gamma_{m_{2}}\right) \nonumber\\
&  \times\frac{\cosh\left[  \pi\gamma_{m_{2}}\left(  1-2\left\vert
x_{1}\right\vert \right)  \right]  }{\sinh\left(  \pi\gamma_{m_{2}}\right)
}\exp\left(  2\pi im_{2}x_{2}\right)  ,
\end{align}
and $M_{2}$ from Eq. (\ref{a44}) and (\ref{lx}) turns out to be just $L\left(
x_{1},x_{2}\right)  $%
\begin{equation}
M_{2}=L\left(  x_{1},x_{2}\right)  .
\end{equation}
Combing $H_{2},$ $J_{2}$ and $M_{2}$ we obtain the form of $G_{2}$ given in
Eq. (\ref{a35}). Considering finally the case for $d>2$ case, we can obtain
$G_{d}$ again from Eq. (\ref{mkh}). Now $K_{d}$ and $H_{d}$ are convergent and
$M_{d}$ can be obtained using the recursive relation Eq. (\ref{a50}). For
example:%
\begin{align}
\left(  M_{3}-\frac{1}{\left(  x_{1}^{2}+r_{\perp}^{2}\right)  ^{1/2}}\right)
&  =\frac{C_{2}}{C_{1}}\left[  M_{2}+\ln\left(  r_{\perp}\right)  \right]
+2\frac{C_{2}}{\left(  2\pi\right)  ^{1/2}}\sum_{m_{2}=1}^{\infty}\cos\left(
2\pi m_{2}x_{2}\right) \nonumber\\
&  \times\sum_{m_{3}}^{\prime}K_{0}\left(  2\pi m_{2}\sqrt{x_{1}^{2}+\left(
m_{3}-x_{3}\right)  ^{2}}\right) \nonumber\\
&  +\sum_{k=1}^{N-1}\left(  \frac{1}{\left[  \left(  x_{2}+k\right)
^{2}+r_{\perp}^{2}\right]  ^{1/2}}+\frac{1}{\left[  \left(  x_{2}-k\right)
^{2}+r_{\perp}^{2}\right]  ^{1/2}}\right) \nonumber\\
&  -2\gamma-\left[  \psi(N+x_{2})+\psi(N-x_{2})\right] \nonumber\\
&  +\sum_{l=1}^{\infty}\binom{-\nu}{l}r_{\perp}^{2l}\left[  \zeta\left(
2l+2\nu,N+x_{2}\right)  +\zeta\left(  2l+2\nu,N-x_{2}\right)  \right]  ,
\end{align}
where $M_{2}$ has already been evaluated above. We see that in all the case,
expression could be written in a form that the essential Coulomb singularity
as the two charges approach each other has been removed.

\section{Conclusions}

Using the limiting behavior of the modified Bessel functions, we showed how
conditionally convergent Coulomb sums may be handled in an elegant way. We
gave two representations of the Green's function for the Poisson equation in
any integer dimensional space. A recursive method was derived that can be
applied for wholly periodic cases, as well as for those cases where one may
have open boundary conditions along one of the directions. The method may be
extended to cover the case where any number of directions may be open. The
formulas obtained show rapid convergence in all part of the simulation cell.
This method is general enough that it can be easily generalized for a higher
dimensional \textquotedblleft triclinic\textquotedblright\ cell. A particular
case of the application of this method for a triclinic cell can be seen in a
recent paper\cite{tyagijcp}. We have shown that the present work generalizes
the work of several authors on periodic and partial periodic systems
\cite{lekner,sperb,mazar}. To our knowledge, this treatment is the first of
its kind ever taken in a dimension higher than $d=3.$

\end{document}